\documentstyle[preprint,revtex]{aps}
\begin{document}
\draft
\preprint{talk presented at STATPHY19, China, 1995}
\begin{title}
Exactly solvable extended Hubbard model
\end{title}
\author{D. F. Wang}
\begin{instit}
Institut de Physique Th\'eorique\\
Ecole Polytechnique F\'ed\'erale de Lausanne\\
PHB-Ecublens, CH-1015 Lausanne, Switzerland
\end{instit}
%%%%%%%%%%%%%%%%%%%%%%%%%%%%%%%%%%%%%%%%%%%%%%%%%%%%%%%%%%%%%%%%%%%%%%%
\begin{abstract}
One dimensional chiral Hubbard model reduces to the Haldane-Shastry
spin chain at half-filling with large but finite on-site energy $U$.
In this talk, we show that the Gutzwiller-Jastrow wavefunctions
are the eigen-states of the Hubbard model at $U=+\infty$
at less than half-filling. The full energy spectrum and
an infinite set of mutually commuting constants of motion
are also given in this limit for the system.
\end{abstract}
%%%%%%%%%%%%%%%%%%%%%%%%%%%%%%%%%%%%%%%%%%%%%%%%%%%%%%%%%%%%%%%%%%%%%%%%
Low dimensional electron systems of strong correlation have been
of considerable recent interests, particularly due to the
discoveries of high $T_c$ materials and quantum Hall effects.
For quantum systems of many degrees of freedom, usually it is impossible
to solve the equations of motion exactly, because of many degrees of
freedom and the interactions of the particles.
However, in some special cases, where systems have infinite
number of simultaneous constants of motion, the systems are
completely integrable. Exact solutions have provided us with
an interesting {\it nonperturbative} approach to strongly correlated
electron systems. The field of solvabilities has attracted
attention from mathematicians, high energy theorists, as well as
from condensed matter theorists. Interesting condensed matter
models can be solved with Bethe-ansatz, such as the one dimensional
Hubbard model\cite{r:wu,r:yang} (a unique example of Luttinger liquid),
the Kondo model\cite{r:andrei} (a local Fermi liquid),
the Anderson model\cite{r:wieg}
for moment formation, and so on.

In this talk, we review some recent results on the one dimensional
chiral Hubbard model. This system was introduced a few years ago
by Gebhard and Ruckenstein\cite{r:Rucken91}, and it
was conjectured to be integrable
for any on-site energy. One very interesting aspect of
this system is that the Mott-Hubbard transition occurs at
non-zero on-site energy. At half-filling,
the Hubbard model in the large on-site energy
limit reduces to the Haldane-Shastry spin chain
\cite{r:haldane,r:shastry}. By far, the complete integrability
is still lacking for finite $U$. The special case of strong interaction
$U=+\infty$ has been studied rather completely, with explicit construction
of wavefunctions, analytical derivation of full energy spectrum,
thermodynamics, as well as the proof of
integrability\cite{r:wang1,r:wang2,r:wang3}.

The Hamiltonian for the one-dimensional Hubbard model is given by
\begin{equation}
H=\sum_{i\ne j;\sigma=\uparrow,\downarrow} t_{ij} c_{i\sigma}^{\dagger}
c_{j\sigma} + U \sum_{i}
n_{i\uparrow} n_{i\downarrow},
\label{eq:hamil}
\end{equation}
where $c_{i\sigma}^{\dagger}$ and $c_{i\sigma}$ are creation and
annihilation operators at site $i$ with spin component $\sigma$.
We take $t_{ij}=it (-1)^{(i-j)}
/d(i-j)$ where $d(n)={L\over\pi}\sin(n\pi/L)$
is the chord distance\cite{r:Rucken91}.
Here we assume periodic boundary condition for the wavefunctions
for odd $L$, or anti-periodic boundary condition for even $L$.

In the strong interaction limit $U=\infty$, each site can
be occupied by at most one electron.
In the following, where-ever in case of the strong interaction, we always
implicitly assume no double occupancy. Let us denote the number of holes
by $Q$, the number of down-spins by $M$.
Following notations used in previous literatures,
the state vectors can be represented by creating spin and
charge excitations from the
fully polarized up-spin state $|P>$,
\begin{equation}
|\Phi> = \sum_{\alpha, j} \Phi (\{x_{\alpha}\},\{y_j\})
\prod_{\alpha} b_{x_\alpha}^{\dagger}
\prod_{j} h_{y_j}^\dagger |P>,
\end{equation}
where $b_{x_\alpha}^{\dagger} =c_{x_\alpha\downarrow}^{\dagger}
c_{x_\alpha\uparrow}$ is the operator to create a down-spin at site
$\alpha$, and
$h_{y_j}^{\dagger} = c_{y_j\uparrow}$ creates a hole at site $j$.

To describe uniform motion and magnetization, consider the
following generalized Gutzwiller-Jastrow wavefunctions\cite{r:wang1},
\begin{eqnarray}
&&\Phi (x, y; J_s, J_h) = \exp {2\pi i\over L} (J_s\sum_{\alpha}
x_{\alpha} +J_h \sum_i y_i) \times \Phi_0\nonumber\\
&&\Phi_0 =\prod_{\alpha<\beta} d^2(x_\alpha -x_\beta)
\cdot \prod_{\alpha i} d(x_\alpha -y_i) \cdot \prod_{i<j}
d(y_i-y_j),
\end{eqnarray}
where the function $d(n)$ is defined as before.
The quantum numbers $J_s$ and $J_h$ govern the momenta of
the down-spins and holes, respectively. They can be integers or half
integers so that we have appropriate periodicities (or
anti-periodicities)
for the wavefunctions under the translations
$x_\alpha \rightarrow x_\alpha +L$, or $y_i \rightarrow y_i +L$ for
odd $L$ (or even $L$).

These Jastrow wavefunctions have been shown to be
eigenstates of the Hamiltonian,
with eigenenergies given by
\begin{equation}
E(J_s,J_h)=(2\pi  t/L) [2 J_h -J_s +L/2] Q,
\end{equation}
where the momenta of the holes and down-spins meet the following conditions:
\begin{eqnarray}
|J_h| \le L/2 -(M+Q)/2,\nonumber\\
|J_h-J_s+L/2 | \le M/2.
\end{eqnarray}

For other excitations, one may assume following more generalized Jastrow
functions:
\begin{equation}
\phi = \phi_s (X, Y) \phi_h (Y) \times \Phi_0.
\end{equation}
Here the functions $\phi_s$ and $ \phi_h$ are polynomials of
$X=\{\exp(2\pi i x_{\alpha}/ L)\}, Y=\{\exp (2\pi i y_i/ L)\}$.
They are totally symmetric
in their arguments, respectively.
The eigen-energy equation thus reduces to
\begin{equation}
(2\pi t /L) [\sum_{i=1}^Q \partial_i (\phi_s \phi_h)
+ \sum_{i=1}^Q \phi_s (\partial_i+L/2) \phi_h] = E \phi_s \phi_h,
\end{equation}
where $\partial_i = Y_i\partial/\partial Y_i$. This eigen-value equation
can be solved exactly, yielding the spectrum given by
\begin{equation}
E =(2\pi t /L) [  \sum_{i=1}^Q n_i + \sum_{\mu = 1}^{Q}
m_{\mu}].
\end{equation}
The integers (or half integers ) satisfy the conditions
$|n_i| \le M/2, |m_{\mu} |
\le L/2 -(M+Q)/2$, where $n_i \le n_{i+1}$ and $m_{\mu} \le m_{\mu +1}$.
This result shows that the spectrum is invariant
when changing the sign of $t$.

In terms of a set of conjugate quantum numbers $K_i = n_i +m_i +(L-Q)/2 +i$,
the spectrum can be rewritten as
\begin{equation}
E = - (2\pi t /L) \sum_{i=1}^Q K_i
+ (\pi t Q /L) (L+1),
\end{equation}
where $ K_i$ takes values from $(1, 2, \cdots, L)$.
One regards these quantum numbers as
the momenta of quasi-particles ``holons''. This gives the full energy
spectrum of the system.
The unoccupied numbers can be considered as
the momenta for the quasi-particles of spin degrees.
Our numerical result shows that the degeneracy of each energy level
is given by the number of the ways to
distribute $L-Q-M$ spins $ s=+{1 \over 2}$ and $M$ spins
$s=-{1\over 2}$
among the empty values.
For fixed number of electrons $N_e = L-Q$ on the lattice,
the free energy consists of two parts,
$F=F_1 - T N_e \ln 2$, where the second term comes
from the decoupled spin degrees of freedom, and $F_1$ is the contribution
from the charge degree of freedom, which is that of
$Q$ spinless fermions with the relativistic spectrum. We see that
the spins and charges are completely decoupled in the free energy,
and this system is a simple example of Luttinger liquid.

Finally, I wish to note that the integrability can be proved
analytically\cite{r:wang3}. All the results for the $SU(2)$ case can be
generalized
to the $SU(N)$ case in this strong interaction limit\cite{r:wang2}.
Recently, some exact eigenfunctions for this model at finite $U$
have been provided\cite{r:shen}. It is hoped that further analytical results
on finite $U$ will be found.

I would like to thank Mo-lin Ge and F. Y. Wu for
the invitation. This work was supported in part by the Swiss
National Science Foundation. Part of the work presented here
was done when I was at Princeton with P. Coleman, to whom
I have been grateful. I also wish to thank Ch. Gruber for
the collaboration at the IPT-EPFL.

%%%%%%%%%%%%%%%%%%%%%%%%%%%%%%%%%%%%%%%%%%%%%%%%%%%%%%%%%%%%%%%%%%%%%%%%%%%

\end{document}